\documentclass[aip,apl,twocolumn,amsmath,amssymb]{revtex4}  
\usepackage{graphicx}
\usepackage{dcolumn}   
\usepackage{bm}        
\usepackage{amssymb}   
\usepackage{epsfig}
\usepackage{amsmath}
\usepackage{mathtools}

\usepackage{float}
\usepackage[colorlinks,urlcolor=blue]{hyperref}
\newcommand{\be}[1]{\begin{equation}\label{#1}}  
\newcommand{\ee}{\end{equation}}     
\renewcommand{\vec}[1]{\mathbf{#1}}   
\let\oldhat\hat
\renewcommand{\hat}[1]{\oldhat{\mathbf{#1}}}
\begin{document}   

\title{ A single-lens universal interferometer: towards a class of frugal optical devices }   

\vspace{2.5cm}

\author{Pooja Munjal \& Kamal P. Singh } 

\email[Corresponding author:]{kpsingh@iisermohali.ac.in; poojamunjal4@gmail.com}   

\affiliation{ Department of Physical Sciences, IISER Mohali, Sector-81, Manauli 140306, India. }

\begin{abstract} 
The application of precision interferometers is generally restricted to expensive and smooth high-quality surfaces. Here, we offer a route to ultimate miniaturization of interferometer by integrating beam splitter, reference mirror and light collector into a single optical element, an interference lens (iLens), which produces stable high-contrast fringes from \emph{in situ} surface of paper, wood, plastic, rubber, unpolished metal, human skin, etc. A self-referencing real-time precision of sub-20 picometer ($\sim \lambda/30000$) is demonstrated with simple intensity detection under ambient conditions. The principle of iLens interferometry has been exploited to build a variety of compact devices, such as a paper-based optical pico-balance, having 1000 times higher sensitivity and speed, when compared with a high-end seven-digit electronic balance. Furthermore, we used cloth, paper, polymer-films to readily construct broadband acoustic sensors possessing matched or higher sensitivity when compared with piezo and electromagnetic sensors. Our work opens path for affordable yet ultra-precise frugal photonic devices and universal micro-interferometers for imaging.

\end{abstract}

\pacs{}
\maketitle

Over the past centuries, a wide variety of ultra-precise optical interferometers have been designed and employed for fundamental and technological applications \cite{michelson,fabry,mirau,mach-zehnder, zygo}, including detections of gravitational waves with Michelson and Fabry-Perot interferometers \cite{ligo, Kenneth}. The precision in these interferometers comes at a cost of multiple high-quality components such as beam splitter, reference mirror, collimator lenses, and optical quality test-surfaces.   
In addition, most interferometers demand tedious alignment and stabilization of its multiple components against non-fundamental mechanical and thermal noises \cite{zygo}. Devising a compact interferometer capable of achieving real-time picometer scale precision by direct interferometry on arbitrary surfaces remains a challenge.

Although, several promising optical techniques have been devised to probe rough or nano-patterned surfaces, for example, electronic speckle pattern interferometry\cite{ESPI}, diffuse optical tomography and imaging \cite{DOT}, interferometric scattering microscopy(iSCAT)\cite{iSCAT1,iSCAT2}, and holography \cite{holographic}. The vertical scanning white-light interferometry, based on Mirau or Zygo interferometers, has been successful to probe patterned surfaces by processing white-light fringes \cite{zygo, scatterplate}. The precision and speed in such techniques is limited by pixel size, scanning speed and signal-processing algorithms. With the advancement in frugal science, successful efforts are made to realize low-cost devices \cite{foldoscope, paperfuse, muPAD2}. However, realization of compact optical devices using precision optical techniques has not been achieved yet.   

\begin{figure}[b]
\centering
\includegraphics[width=0.95\columnwidth]{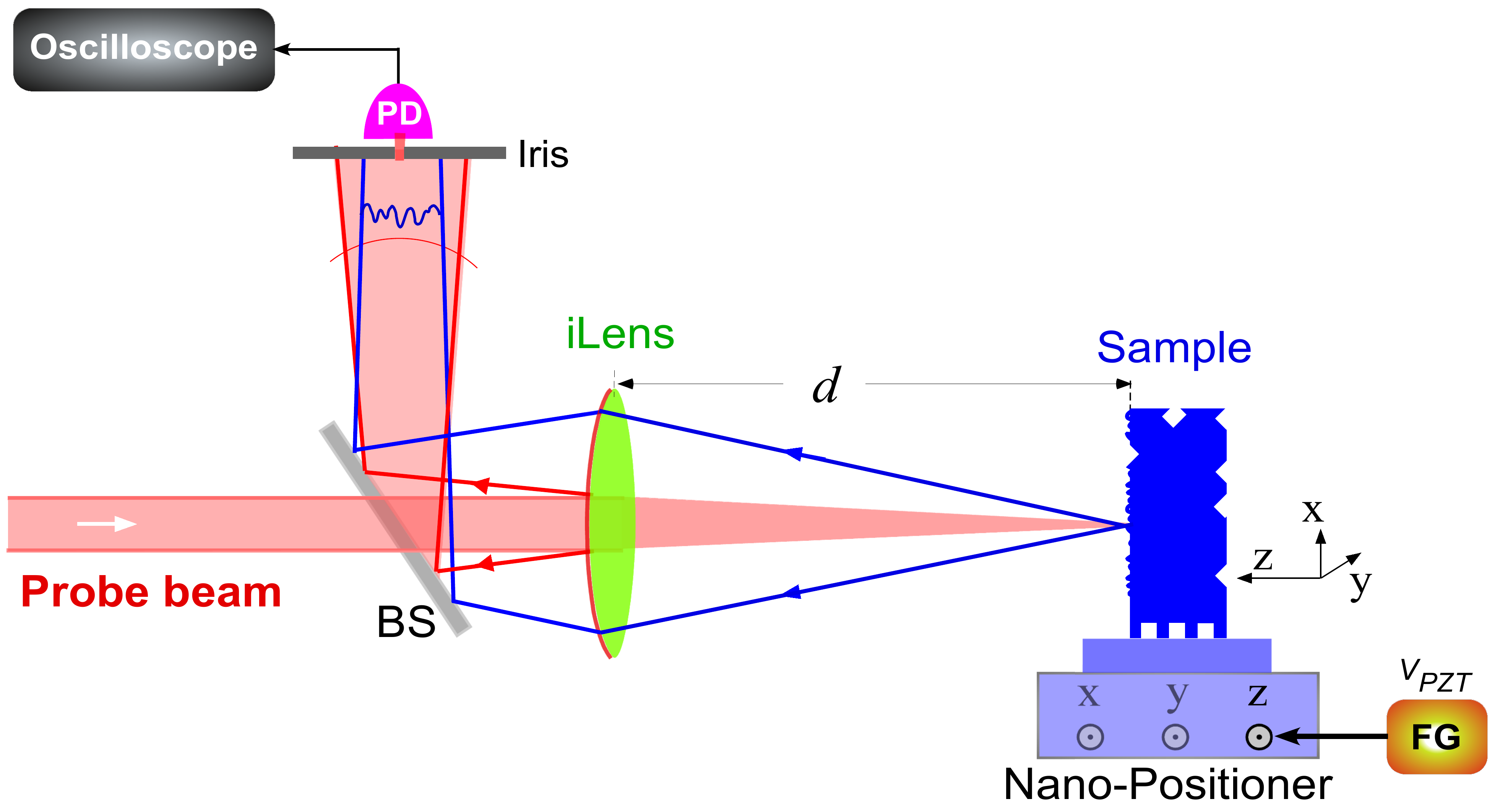}
\caption{Schematics of iLens interferometer. An iLens produces high-contrast structured fringes from a surface using a low-power He-Ne probe laser ($10~mW$, $\lambda = 632~nm$). The z-position of the sample from the iLens, $d(t)$, is controlled by a piezo-positioner driven by $V_{PZT}$ from a function generator (FG). The central intensity of fringes is detected with a photodiode (PD).}

\end{figure} 

Here, we offer a route to ultimate miniaturization of interferometers by integrating essential functions of beam splitter, reference mirror and light collection optics into a single-lens enabling stable, high-contrast fringes with various surfaces including paper, plastic, wood, metal, rubber, human skin, hair. Our universal interferometer achieves a real-time precision of sub-$20~pm$ ($\sim \lambda/30000$) with a simple detection of interference intensity using a high-speed photodiode. This approach allows us to construct a wide variety of interferometric devices such as, a paper-based picogram weighing balance and frugal acoustic sensors which match or outperform their commercial counterparts.

Fig.~1 shows the schematic of our experimental set-up. The key optical element of our interferometer is a convex iLens whose front surface was partially silver-coated. It is a three-in-one component working as beamsplitter, reference mirror and light collector. It splits a fraction of incoming probe beam, generates a reference beam and also collects weak diffuse reflection from the sample surface kept at a distance $d$. Both these beams are superimposed collinearly to produce a high-contrast interference pattern on the screen. In effect, the iLens with any test surface forms a single-lens interferometer. 

The iLens interferometer produced high-contrast stable fringes from a wide variety of common surfaces. 
Fig.~2(a-f) shows six representative examples of interference pattern from cloth, human skin, cellulose paper, quartz crystal, silicon rubber, and Ag micro-wire. Additional data on 20 samples covering most of the readily available materials and micro-objects ($2-50 ~\mu m$ diameter) are shown in Supplementary Fig.~S1. The fringe shape was determined by sample properties such as surface roughness and curvature \cite{born00}. However, our precision is achieved for all the cases since, the interference contrast $C=(I_{max}-I_{min})/(I_{max}+I_{min})$, ($I_{max}$ and $I_{min}$ are maximum and minimum light intensity, respectively) was $70-95\%$ for all the samples. The contrast can be further optimized using appropriate coatings (see Supplementary Table~S1). 
To measure self-calibrated picometer measurements, we record interference intensity of the central fringe with an iris and PD. {\bf The iris opening was sub-mm diameter and the PD sensor was $800~\mu m^2$ which are adjusted much smaller than typical fringe-size}.

\begin{figure}[t]
\centering
\includegraphics[width=0.9\columnwidth]{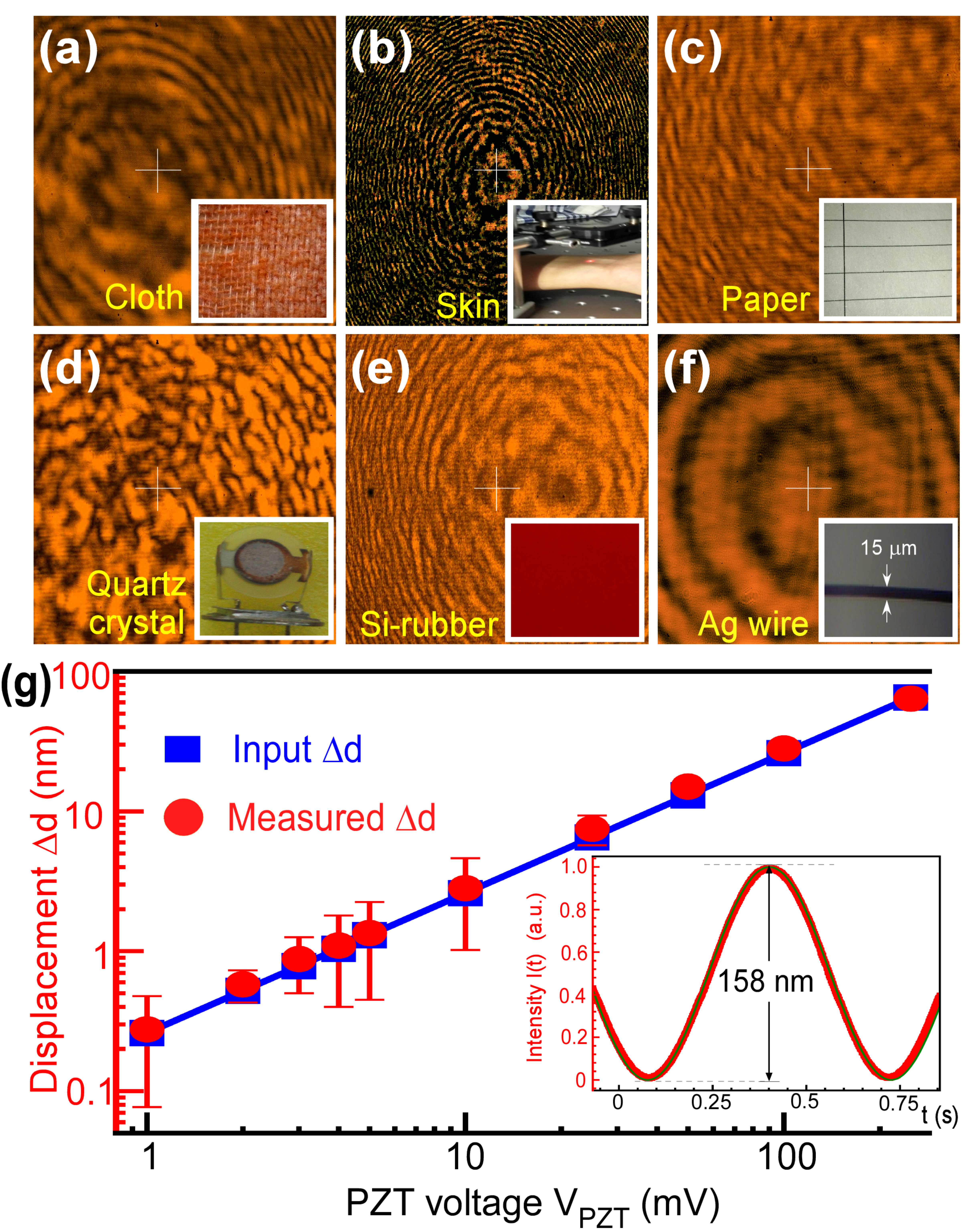}
\caption{Snapshots of fringes and Self-calibrated picometer measurement.
(a-f), Snapshot of the fringe with picture of sample in the inset. The interference was obtained with the iLens, $f=5~cm$, $R=0.4$ and $d \approx 55~mm$.
(g), Calibration curve showing input sample displacement (blue square) with the one independently measured using PD (red circle). Solid line is a linear fit. Error bars indicate experimental noise-floor. Lower inset shows the normalized fringe intensity $I(t)$ at cross-hair when the lens-to-sample distance $d$ is scanned by driving piezo-positioner.
}
\end{figure}

\begin{figure*}[t]
\centering
\includegraphics[width=1.5\columnwidth]{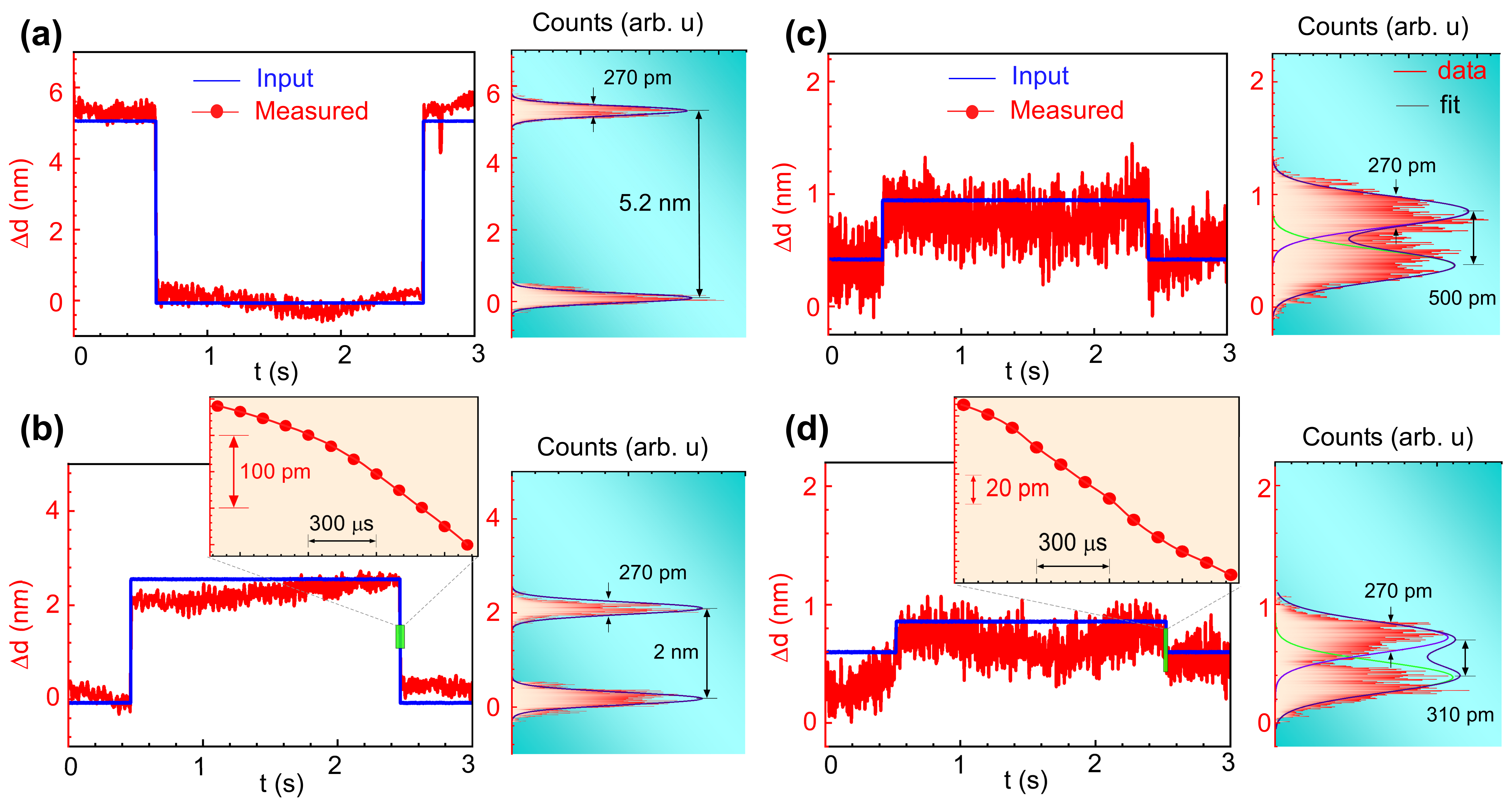}
\caption{Real-time picometer resolved surface dynamics. (a-d) Measured displacement and input displacement versus time for cardboard surface at different amplitudes (a) $5.2~nm$, (b) $2.0~nm$, (c) $500~pm$ and (d) $310~pm$. The figures on the right-side of each graph indicate histogram of signal indicating the noise-floor and resolution between two levels. Insets: zoom showing sub-$20~pm$ change in $\Delta d$ within $300 \mu s$ interval. Histograms fit have Gaussian distribution with standard deviation of $270~pm$.}
\end{figure*}

A simplified analysis below illustrates self-calibration of our interferometer. {\bf The reference beam propagating along the $z$ axis and produced by iLens on the screen $(x,y)$} is given as $\vec{E_R} = E_{R0}(x,y) e^{\iota (k z_1 -\omega t)} $, with $k=2 \pi/\lambda$. The diffuse reflection from the sample on the screen can be written as \cite{born00},  
        $\vec{E_S} = E_{S0}(x,y) e^{\iota (k z_2 -\omega t + \xi(x,y))} $, where $z_1$ and $z_2$ are optical path lengths from the iLens and the sample, respectively. $\xi(x,y)$ denotes a random phase difference between the two waves on the screen. The total intensity $I(x,y)$ is given by,
        $I(x,y) \sim \vert \vec{E_R}$ + $\vec{E_S} \vert ^2$. Considering that the peak amplitudes of the reference and the scattered beams are identical, on solving and rearranging the terms we get,     
        $I(x,y) = I_0 \cos^2[\frac{1}{2}(\beta (z_1-z_2) + \xi(x,y))]$, with $\beta= 4 \pi/\lambda$ and $I_0$ maximum intensity. The first term in the argument is deterministic, whereas, the second one is a random term. {\bf Since we experimentally measure the central intensity through iris $I(x_0,y_0)$ at ($x_0,y_0$) along the interferometer axis the plane-wave approximation is well-justified and $\xi(x_0,y_0)$ becomes a constant.} The local intensity then follows the Michelson-like dependence  \cite{pm-michelson, dogariu}, $I(x_0,y_0,t) \sim cos^2(\frac{1}{2}\beta d(t))$, as $d(t)= z_1-z_2(t)$ is varied.

We established self-calibration of our set-up for picometer measurements \cite{cal}. 
For this, we provided an input displacement $\Delta d = d(t)-d(0)$ to the sample using the nanopositioner and compared the displacement measured from the PD signal (see Fig.~S2(a,b)"Multimedia View" for dynamic fringes).  
The input and measured displacements agreed very well over the entire measurement range covering three orders of magnitude (Fig.~2g). Similar data validating our calibration procedure on four different rough samples is shown in Fig.~S3. The actual resolution in our displacement measurement is sub-$20~pm$, as we will show in Fig.~3, however, lack of reliable method for controlling displacement of rough-sample below $266~pm$ determined the lowest data point in the calibration curves (Fig.~2(g), Fig.~S3). We emphasize that the self-calibration must rely on the interference condition, i.e., half-fringe collapse corresponds to $\Delta d = \lambda/4 \approx 158~nm$, and it is independent of the time-scale of the fringe collapse.

In Fig.~3(a-d), we demonstrate the use of interference intensity for real-time sub-20~pm precision in measuring dynamic displacement of arbitrary surface, such as a cardboard. The sample was imparted a single input displacement-pulse (in blue) of different peak amplitudes $5.2~nm$, $2.0~nm$, $500~pm$ and $310~pm$ and resulting displacement was independently measured (in red) with a PD and oscilloscope. The measured displacement $\Delta d$ agreed very-well with the input. On the right-hand side of each curve, we show the histograms of noise showing a standard deviation of about $270~pm$, {\bf resulting from residual mechanical, beam pointing and detection electronics noises}. The zooms near steps in the insets of Fig.~3(c,d), show our ability to resolve sub-20~pm displacements within $300~\mu s$. Similar data for other samples are presented in Fig.~S4 to prove repeatability in our measurements. {\bf The dynamic range of displacement for our interfermeter is $2 Z_R/\Delta d$ where $Z_R$ is Rayleigh range of iLens and $\Delta d =20~pm$ was around $10^9$ for near focus operation.}

It is worth mentioning that the demonstrated picolevel sensitivity was achieved with simple real-time detection of interference intensity with a PD, without lock-in, heterodyne, or noise-filtering techniques \cite{ligo, michelson}. Such advanced techniques may further enhance resolution of our interferometer close to fundamental thermal-noise limit albeit for rough surfaces. Next, we will show a breadth of application possibilities of iLens interferometry in constructing various ultra-precise devices from common materials. 

We fabricated a paper-based optical balance enabling rapid picogram precision measurements. 
We folded a strip of paper ($80~\mu m$ thick, 4~mm wide, L=1.5~cm) in zigzag-pattern to design an elastic load-cell \cite{paper-folding}, as shown in Fig.~4(a,b). A weight force on the middle of the load-cell produced elastic deformation $\delta y$, that was directly measured with $pm$ precision with iLens to estimate unknown mass $m=(k/g) \delta y$, where $g$ is the acceleration due to gravity, and $k$ is the spring-constant of the load-cell. 
We measured the spring-constant of the paper-load cell by pulling test which validated its Hookean response with $k=24~N/m$ (see Supplementary Fig.~S5) upto a maximum load limit of about $0.6~mN$. 

\begin{figure}
\centering
\includegraphics[width=0.9\columnwidth]{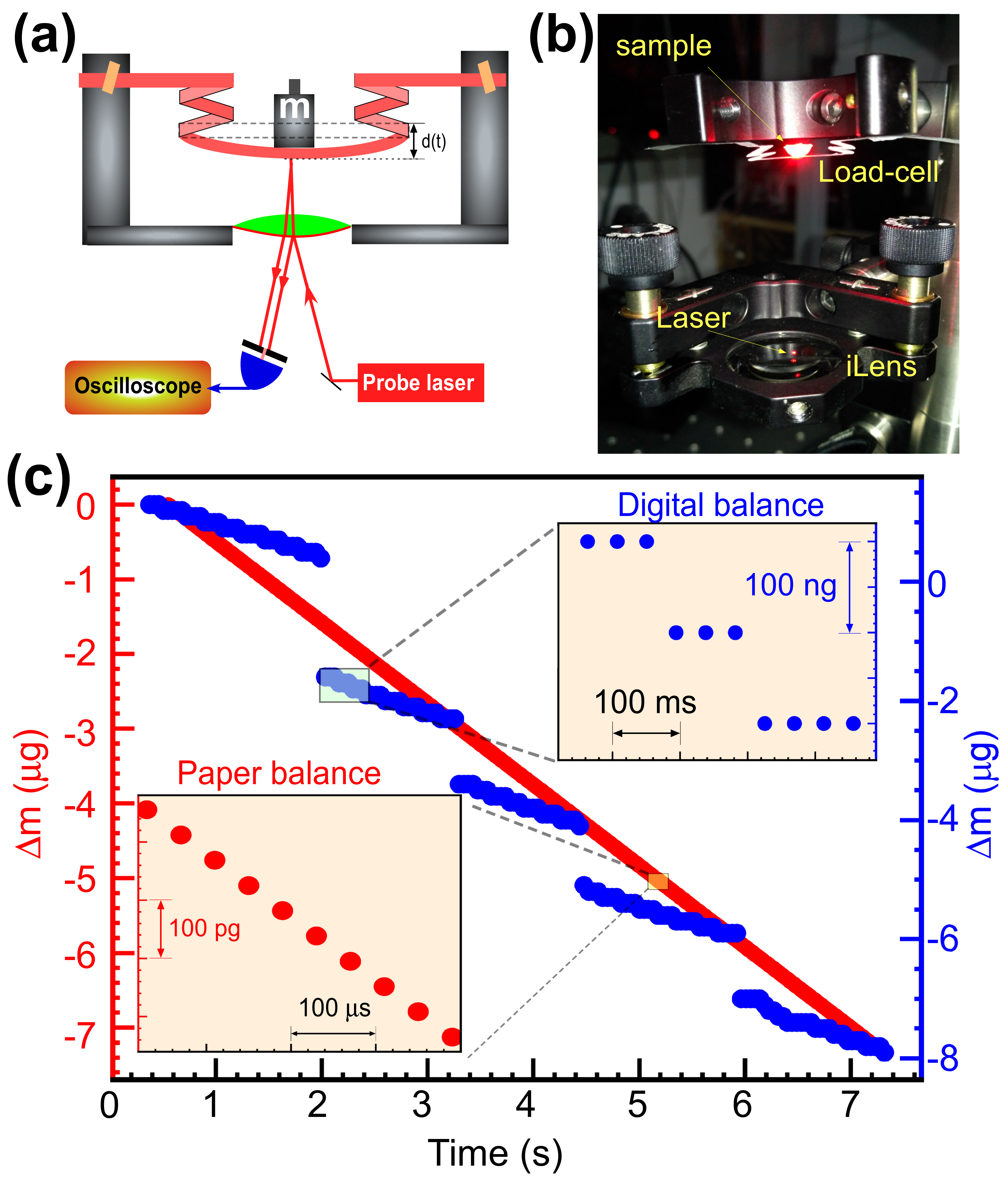}
\caption{Paper-based pico balance. (a), Schematic of the paper-load cell based optical balance. (b), A labelled picture of the weighing set-up. (c), Comparison of weight-loss of an evaporating water drop with paper balance (red data) and electronic ultra-micro balance (blue data). Zoom of the electronic balance data shows $100~n g$ resolution. Zoom of the paper-balance shows about $100~pg$ with 1000 times faster readout. Error bars are about the size of the data symbols. }
\end{figure}

The weighing performance of our paper-balance was compared with a commercial electronic ultra-microbalance which is based on the electromagnetic force compensation technology \cite{EFCbalance}. We measured, for example, mass of an evaporating water drop under identical conditions of temperature, humidity, volume and surface area. As shown in Fig.~4(c), the ultra-microbalance exhibited discrete weight-jumps corresponding to $100~ng$ precision (one part in ten million) within tens of ms. In contrast, our paper-balance is 1000 times more precise with sub-$100~pg$ weighing accuracy. Moreover, thanks to the direct optical measurement of the load-induced displacement, our device can provide 1000 times faster measurements.  

Previously, the precision weighing in femto-to-zepto grams range was achieved using nano-mechanical resonators by measuring the load-induced shifts in its resonance frequency \cite{zepto-gm-Xe}. Such nano-beams are carefully fabricated with dedicated nano-processing of SiN and require high-vacuum and cryogenic conditions to achieve zepto gram precision. Various micro-cantilevers have also been designed to weigh living cells, tissues with femto-gram precision \cite{DNA-zeptogm, weighingCell}. Even though, our balance is compact, and easy to fabricate and calibrate, it can achieve picogram precision in ambient conditions.

\begin{figure}[t]
\centering
\includegraphics[width=0.9\columnwidth]{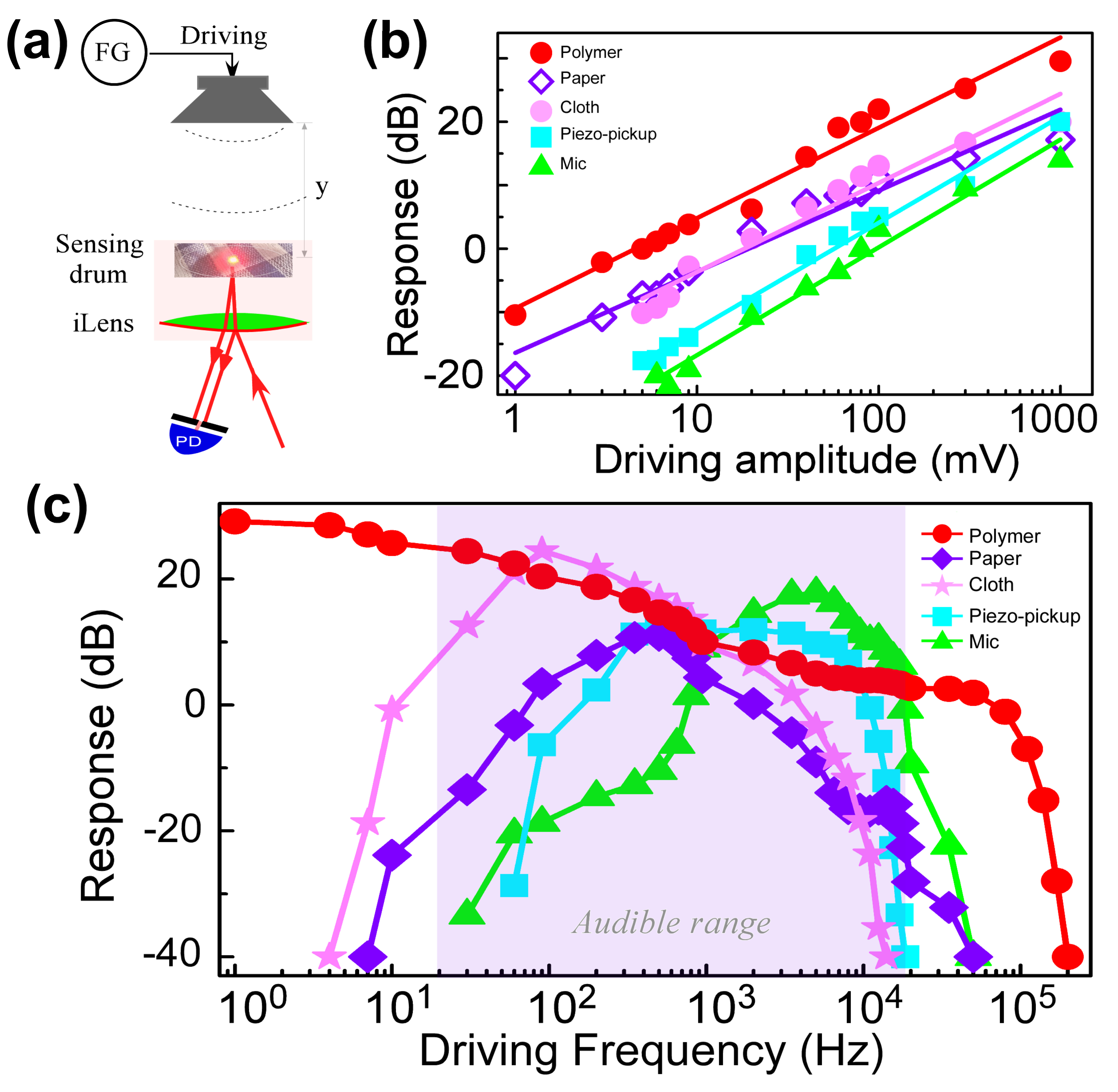}
\caption{Design and characterization of frugal optical acoustic sensors. (a) Schematic of iLens acoustic sensor made from cloth/paper/polymer, (b), Comparison of detection sensitivity of frugal sensors with piezo and microphone at $1000~Hz$. Source amplitude was varied by driving voltage for a fixed source-sensor distance. (c), Frequency response of iLens acoustic sensors when compared to piezo-pickup and microphone. Response (dB) is defined as $20Log_{10}(V_{sensor}/V_{ref})$ at $V_{ref}=1~mV$.}
\end{figure}

In another application, we used cloth, paper or polymer membrane as an ultra-sensitive acoustic sensor operating from $1~Hz$ to $200~kHz$. We obtained interference pattern from such diaphragms and used interference intensity, as described previously in Fig.~2, to measure its dynamic displacement in real-time. Three example devices were fabricated from the three common materials, a cotton cloth (around $200~\mu m$ mesh size, diameter $D=12.5~mm$), a paper-load-cell, and a polymer diaphragm film. We validated our devices by subjecting it with acoustic driving at different frequencies and different driving amplitudes, as shown schematically in Fig.~5(a).  

It is worth comparing the bandwidth and sensitivity of our acoustic devices with easily available commercial sensors such as piezo-pickup and an electromagnetic microphone. For this, all the devices under test were kept at a fixed distance ($y=5~cm$) from an acoustic source, and their response to acoustic signal was detected from 1~Hz to 200~kHz driving frequency. As shown in Fig.~5(c), our custom-built sensors exhibited broadband response from infra-sound to near-ultrasound frequencies, while the piezo and electromagnetic sensors responded mostly in the audible range. All these sensors operated in the linear-regime for their comparison. Besides, owing to our interferometric read-out our sensors exhibited higher sensitivity (e.g., at $1~kHz$) to weak driving amplitude of the source (Fig. 5(b)). We attribute the maximum cut-off frequency of our devices to the inertia of the vibrating drum since we drive them far beyond their natural resonance frequency \cite{momentum}. In fact, when the vibrating polymer diaphragm of a commercial headphone speaker was read with the iLens, instead of electrically, we could achieve enhanced bandwidth as well as sensitivity (compare curves in red and green in Fig.~5(b,c)). Our frugal acoustic devices present an attractive alternative for various applications demanding low-cost precise displacement sensing, such as in medical instruments \cite{acousticForce, MechBiosensor} and photo-acoustic spectroscopy for trace-gas detection \cite{gasera}.

Our interferometer offers advantages over conventional Michelson \cite{michelson}, Mirau \cite{mirau} or Fizeau interferometers \cite{zygo1} which essentially require precise stabilization and alignment of its multiple components like beam-splitter, reference mirror and light collector. 
{\bf We experimentally observed that the iLens interferometer is robust with tilt misalignment of the iLens ($\pm10^\circ$) and has flexible working distance within several $Z_R$.}  
Importantly, its unique simple design enables precision interferometry with any surface like no other existing interferometers including the stimulated micro-objects.
 
It should be possible to further enhance the precision in femtometer regime, approaching the fundamental noise limit, by employing frequency stabilized lasers and advanced signal detection such as lock-in and heterodyne techniques \cite{pm-michelson, ligo}. In addition, using high-speed avalanche detectors one should achieve sub-$ns$ time-resolution. In applications demanding compactness, one can design a micro-iLens with micro-lasers and detectors on 3d printed substrate.   

In summary, we demonstrate a paradigm of universal interferometry with everyday materials enabling self-calibrated real-time measurements with picometer resolution. Using multifunctional iLens we demonstrated a variety of precision devices including a paper based picobalance and frugal yet ultrasensitive acoustic sensors.

We envision diverse applications of our technique in probing complex fluids, pristine biomaterials, and plasma with unprecedented precision. Our compact and easy approach will be attractive in monitoring atomic level stability of satellites and space-borne instruments, in biomedical engineering, in 3d printing of interferometer based research equipments and for high-quality education in low-resource settings. Employing twisted light in our {\bf iLens interferometer may open intriguing opportunities for probing picometer-resolved complex light-matter interaction with compact setup \cite{ momentum, spiral, OAM, OAM1}}.   

\vspace{0.5cm}
See {\bf Supplementary Materials} for more data on 20 different samples, evolution of interference fringes and design of iLens coating and additional measurements on paper-based load cell devices.

\begin{acknowledgements}
We acknowledge B. Panda, N. Singh, M. S. Sidhu, V. Vashisht, G. Verma for discussions. PM acknowledges A. Singh for help with Python script. KPS acknowledges financial support from DST India and Max Planck Society Germany.
\end{acknowledgements}

\end{document}